# On Possibility of Electromagnetic Nature of Atmospheric Intensive Vortices Generation

M.D. SCHERBIN

*Federal State Institution
"Russian Federation Ministry of Defense,
12 Central Scientific Research Institute"*

___________________________

* Corresponding Author Address: M.D. SCHERBIN, Federal State Institution "Russian Federation Ministry of Defense, 12 Central Scientific Research Institute", 141307, Sergiev Posad. E-mail: mdsh@spnet.ru

**ABSTRACT**

Our goal is to find out possible appearance of rotational motion in the atmosphere, created by a lightning. A positive solution of this question, in the end, leads to a necessity of accepting that processes of birth and annihilation of electron-positron pairs take place in the leader head of lightning channel. Under this condition, within the framework of our physical model of development of a lightning, we suggest a logical explanation of the stepwise structure of a negative leader, intensive radio emission produced by lightnings, gamma-ray flashes from thunderstorm and anomalously intensive electric currents from positive leaders.  As for the development of intensive tornado it is determined by pumping of the angular moment produced by lightnings in a local area of the thunderstorm.

**Index Terms**: 3324 Atmospheric Processes: Lightning; 3304 Atmospheric Processes: Atmospheric electricity

## 1. Introduction

The interest to studies of storm phenomena in the atmosphere has lately risen significantly. This problem is of a great importance of its own. Besides, a correlation of lightning activity with appearance of destructive tornadoes has been noted [*Carey et al.,* 2003].

In the work of *Ziegler et al.* [2003] possible influence of the lightning activity on tornado appearance is observed. The heightened interest to the afore mentioned problem in this particular direction is connected to lack of consistently established mechanisms of generation of vorticity flows in the atmosphere, leading to appearance of devastating tornadoes. The revealing of such mechanisms could be used in theoretical numeric modeling of tornado-genesis. Impressive lightning activity, accompanying tornado development (up to 600-800 flashes a minute, see [*Land and Rutledge*, 2005; *Buechler et al.,* 2002]), clearly demands a search of the nature of its initiation, connected specifically to the lightning activity in the storm cloud.

The possibly decisive influence of the lightning activity upon the development of intensive vortices has been noted long ago. In the work of *Vonnegut* [1960] an analysis of the problem along with possible mechanisms of this influence is given fairly fully. In scientific literature numerous attempts have been made to connect electromagnetic processes to description of tornado-genesis (e.g. [*Boyarevich et al.*, 1985]), but none of these led to the expected results.

The main difficulty in determining the connection of lightning activity with the process of tornado-genesis is related to the fact that up to present time not the appearance of an angular moment of quantity of motion of the thunderstorm in the process of lightning development, which might initiate the birth of an intensive vortices, has not been established The question of possibility of presence of the angular moment in the process of the development of a lightning is quite legitimate, since, in spite of obvious achievements in the studies of physics of lightning development, this problem still contains many white spots awaiting their resolution. We will point out some. The physical process of lightning appearance is not clear [*Uman,* 1969; *Bazelyan and Raizer,* 2001]. The physical nature of a step leader (e.g. [*Uman,* 1969; *Bazelyan and Raizer,* 2001]) is not well understood. There is no final solution of the question about the physical nature of atmospheric luminosity (see [*Kerr*, 1994]) and sources of gamma-ray flashes during storm activity in the atmosphere (e.g. [*Fishman et al.*, 1994; *Torii et al.*, 2002]). The physical nature of recently found gamma-ray flashes during investigation of trigger lightning (see [*Dwyer et al.*, 2004]) is unclear. The physical nature of extremely strong RF radiation from lightning channels is vague, as well as why power of radiation from a negative leader is one order higher than that from a positive leader (see [*Le Vine,* 1980; *Willett et al.*, 1989; *Smith et al.,* 1999; *Thomas et al.*, 2001]). There are plenty of similar questions. The question of finding the angular moment in the atmosphere is possibly connected with the solution of the questions mentioned above. Considering that during initiation of a lightning and its later development a wide range of electromagnetic processes occurs, Feynman's opinion in his lecture [*Feynman et al.*, 1964a] is worth noting "...if there is a magnetic field and some charges, than the field also has the moment of quantity of motion. It appears as soon as the field itself is created." The lightning development is constantly in the stage of the organization of the process. Because of this, the possibility of acquisition by the surrounding environment of the angular moment should be searched in the entire time of the lightning channel development.

## 2. Basic suggestion

We will point out some processes of lightning channel formation, which are necessary for formulation and substantiation of the main hypothesis, concerning necessity of existence of angular moment during lightning development. A sufficiently detailed account of the physics of lightning development in the atmosphere is presented in the books [*Uman,* 1969; *Bazelyan and Raizer,* 2001].

Lightning channel formation occurs due to the leader process. Schematically, the development of a positive leader happens as follows. The streamers caused by the ionization waves are spread from the leader head. These streamers are forming the streamer zone, the cross section size of which is several meters. The leader head moves to the streamer zone, containing old (dead) and new streamers. As a result, around the lightning channel a charged case is formed, the polarity of which is the same as in the streamer zone. The development of a negative leader is characterized by the stepwise manner of its spreading with the average length of a step 30 m. Its essence is as follows (see [*Uman,* 1969; *Bazelyan and Raizer,* 2001]). In front of the streamer zone of a negative leader movement the plasmic body (Schonland`s pilot leader) with a diameter of ~5-6 m is moving, presenting a bi-leader system: negative streamers spread along the path of its movement; a positive streamer spreads towards a negative leader head – as soon as they connect, the negative leader head moves by a jump into the center of the bi-leader system and the process repeats. The weak stepwise fashion of the positive leader development is pointed out by *Bazelyan and Raizer* [2001], *Berger and Vogrlsander* [1966].

We will consider some data from observations in nature. Analyzing photo-images of lightning channels during their close-distance observation *Uman* [1969] noticed that "...in the least bright pictures the channels where slightly (order of centimeters) twisted and turned." The channel radii lie between 1.5-6 cm. The photo-images of these channels are given by *Evans and Walker* [1963, Fig. 2c]. The uniform structure of channel twisting along its length allows us to conclude that it is twisting immediately behind the leader head. The step of the twisting of a channel is close to its diameter. The channel twisting can be caused only by a magnetic force influence. The mentioned effect may be achieved, if the lightning channel may be described as a long solenoid with some surface current. The structure of the magnetic field inside and at the ends of a long solenoid is well known (e.g. [*Feynman et al.*, 1964b]). In Fig. 1, the solution of the problem on changing of magnetic field induction *B* along a long solenoid along its axis (axis *z*) in the end area ($z < 0$ – inside the solenoid; $z > 0$ – outside the solenoid). Here $r_0$ – is the radius of the solenoid. This solution looks as follows:

$$B = \frac{1}{2} B_0 (1 - z / \sqrt{z^2 + r_0^2}), \tag{1}$$

where $B_0$ – is the induction inside the channel-solenoid.

Magnetic current through the end section is equal to the half of the current through the section inside the solenoid. Another half of the current goes mostly through the surface of the solenoid till the distance from the end approximately equals to the diameter of the solenoid. In this area the component of the magnetic field along the radius of the solenoid ($B_r$) differs from zero. When the current $I_0$ passes through the axis of the current channel, the end area must be twisted by the magnetic force $F = I_0 \times B_r$.

Let us consider, what might be the other consequences of the presence of the longitudinal magnetic field in the lightning channel for its development and what might be its physical nature.

### 3. Discussion

Consider a lightning flash of a bipolar structure, which is the conducting channel between leader heads of opposite polarities. Current $I_0$ flows through the channel axis. The expected magnetic induction along the axis, caused by the azimuth current, is *B* and the azimuth current has the density *i* with respect to the unit of channel length.

When the leader heads moves, the magnetic fields of the end areas move, preserving their structure, and, due to the changes of the magnetic currents in the areas the magnetic fields crossed there is electric field $E_\varphi(r)$ induced, where $r$, $\varphi$ – are the coordinates of cylindrical coordinate system. The induced electric field is located in a narrow strip of the area between the planes, perpendicular to the channel axis and encompassing the area of changes of the magnetic field in the end area. Schematically the structure of currents and fields in the end areas of the lightning channel is given in Fig. 2. The direction of the induced current $I_\varphi$ corresponds to the direction of the induced electric field $E_\varphi$.

Let us consider the directions of the action of volume forces brought to life by the interaction of the longitudinal magnetic field with the currents. Here the magnetic field of a direct current is not considered and is not reflected in Fig. 2. The direction of longitudinal magnetic field will not play a crucial role in the subsequent analysis.

It is worth noticing that the emerging volume forces act only upon the end areas of the current-conducting channel – on the areas behind the leader head with their length equal to its diameter (~10 cm). The volume force, defined by the interaction of the component $B_r$ with longitudinal current $I_0$, leads to twisting of the channel in its opposing ends. The volume force, appearing due to the interaction with component $B_r$ with the induction current $I_\varphi$, independently of the polarity of the leader is directed the way it spreads. Here the following is worth of attention. Conductivity in the area of the case of a positive leader depends on positive ions. In the area of the case of a negative leader directly behind its head the situation is different. The streamer zone of the leader head overlaps the end area of the channel (which is not reflected in the picture) and this gives us a reason to believe that the conductivity of the induction current in the area of the case of a negative leader is electronic. Considering that the mobility of electrons is more than two orders higher than the mobility of ions, the volume force, acting upon the end area of the negative leader, is more than two orders higher than the volume force acting upon the end area of a positive leader. If the impulse received by the end area of a negative leader determines the speed of its movement, which surpasses the speed of the movement of a leader head, than the end area will be torn off. Apparently, the Schonland's pilot leader is an end area of the negative leader, torn off along with its head. During this process the positive leader sustains no considerable transformations, though, as has been mentioned before, the presense of a weak stepwiseness is noticed.

Let us estimate possible values of induction of longitudinal magnetic field and inducted electrical fields.

To estimate the value of the magnetic induction along the channel we will use the fact that the induction allows twisting of the channel with a step approximately equal to the diameter of the channel $2r_0$ (~ 0,1 m). The equation of movement of the conducting gas in the channel, the right part of which contains only the magnetic force, is given by *Alfvén and Falthammar* [1963]:

$$\rho \frac{dv}{dt} = jB_r, \qquad (2)$$

where $\rho$ – is the mass density; $v$ – is the media speed; and $j = I_0/\pi r_0^2$ – is the density of the current along the channel. Since the magnetic force permits the movement of the gas around the circle, we will accept that the acceleration equals $\omega^2 r$, where $\omega = 2\pi/t_0$ – is the angular frequency of the angle movement. During the time of one turn of the particle, the leader head must move the distance equal to $2r_0$, i.e. $t_0 = 2r_0/v_L$, where $v_L$ – is the leader head velocity. For the component $B_r$ of induction vector we receive the following equation:

$$B_r = \rho \frac{\pi^3 v_l^2}{I_0} r. \qquad (3)$$

We will assume that in the end area the function $\partial B/\partial z$ is constant and equal to its value at $z = 0$. Under such an approximation, using the results of *Alfvén and Falthammar* [1963], we can estimate the value $B_r$ from the equation

$$div\, B = 0, \tag{4}$$

assuming $\partial B_\varphi / \partial \varphi = 0$. Under the accepted conditions from the equation (4) it follows that:

$$B_r = -\frac{1}{2} r \frac{\partial B}{\partial z}, \tag{5}$$

where $\partial B/\partial z$ is calculated from (1) with $z = 0$. Thus we have:

$$B_r = \frac{1}{4} \frac{r}{r_0} B_0. \tag{6}$$

From (3) and (6) it follows that

$$B_0 = \rho \frac{4\pi^3 v_L^2 r_0}{I_0}. \tag{7}$$

Accepting that $\rho = 0{,}1\rho_0$, where $\rho_0$ – is the density in the atmosphere under normal conditions; $I_0 = 20$ kA – is the value fixed for a negative leader by *Carey et al.,* [2003]; $v_L \approx 10^6$ m/s (see [*Uman,* 1969]), we get $B_0 \approx 3\cdot10^8$ Tl. This value of $B_0$ is required to twist a lightning channel, the parameters of which are described by *Uman* [1969] and *Evans and Walker* [1963]. It is worth pointing out that the lightning channel end area in question is transitional from the leader head to the channel. Here, as has been pointed out by *Bazelyan and Raizer* [2001], the value of the current may be considerably higher than the observed values, but this does not change the matter – the obtained value of $B_0$ appears to be gigantic.

According to the given value of magnetic induction it is possible to calculate the values of the induced electrical field in the plain which is traversed by the leader head in the period of time of the change of the magnetic current $\Delta\Phi$ through the section $S = \pi r_0^2$ from zero to the final value of $B_0 S$.

$E_\varphi$ can be found from the expression below

$$E_\varphi = \frac{1}{2\pi r} \frac{\Delta\Phi}{\Delta t}, \tag{8}$$

where $\Delta t = 4 r_0 / v_L$.

At the previous values of parameters the expression for $E_\varphi$ is

$$E_\varphi \approx 10^{12} \frac{1}{r}, \text{ V/m}. \tag{9}$$

On the surface of the end area, including also the leader head, $E_{\varphi 0} \approx 10^{13}$ V/m.

Such values of $B_0$, $E_{\varphi 0}$ and the surface currents, appropriate for $B_0$, have not been observed and are not realistic for a lightning channel, but may be realistic in the transition area in questions and in the leader head. The value $E_{\varphi 0}$ is notorious, it is close to the critical value of the electric field ($10^{14}$ V/m), under which Dirak's vacuum polarization occurs, accompanied by the birth of electron-positron pairs [*Grib et al.,* 1980]. The subsequent annihilation of the pairs in the presence of third bodies is accompanied by γ-quantum emission.

What kind of atmosphere processes may bring about such critical fields, when it is known that a lightning is initiated at field values ~ 300 kV/m? This question remains open. Apparently, the existence of the found effect is conditioned by the particularities of the behavior of a large amount of charged ice particles, concentrated in large volume, in the electric field. Until this time, the electric structure of hydrometeors is not studied. The difficulty of the problem can be seen from the papers of *Illingworth,* [1985], *Baker and Dash* [1994]. *A. J. Illingworth* [1985] mentions the opinion, given by P.H. Handel's in 1985 in an unpublished work that ice hydrometeors are possibly ferroelectrics and the process of electrization is a polarization catastrophe.

It may be assumed that the charged ice particles in the external field, under conditions not yet known, cause as well vacuum polarization, which, in the end, leads to quantum processes of birth and annihilation of particles, accompanied by optical phenomenon of a lightning flash.

In principle, the occurrence of these processes in the leader head explains logically the existence of vast zones of atmosphere luminosity above thunderstorms with considerable lightning activity (see [*Kerr*, 1994]), observed from satellites gamma-ray flashes from storms (see [*Fishman et al.*, 1994; *Torii et al.*, 2002]), and also observed gamma-ray burst from trigger lightnings [*Dwyer et al.,* 2004].

An analysis of quantum processes in a leader head of a lightning demands a special study. Yet, during this stage it is possible to trace a quality picture of the influence of these processes on the leader development and to give more accurate definition the picture of the lightning development suggested above.

Accepting legitimacy of the processes pointed out, we may assume that the streamer zone of a leader head, observed as a luminous object, is formed not only due to the ionization waves (see [*Uman,* 1969; *Bazelyan and Raizer,* 2001]), but also due to the intensive γ-quantum emission.

The quanta of emission take away a magnetic moment corresponding to the value of the spin which is equal to one. The magnetic moment of the opposite sign acts on the third particles, which move from the leader head into the lightning channel. The magnetic field in the transition area of the channel is created by the non-compensated magnetic moments of separate particles. According to the data given by *Bazelyan and Raizer* [2001] the lightning channel heats to the temperature of the order of 30 kK, which must lead to destruction of the direction of the magnetic moments of separate particles and to destruction of the longitudinal magnetic field of the lightning channel. Then, due to the difference of mobilities of ions and electrons in the zone of the case of the lightning channel, the process of the destruction of the non-compensated magnetic moments must have its particularities for leaders of different polarities.

Qualitatively this can be pictured as follows. It has been mentioned above that in front of a lightning channel of a negative polarity a luminous plasma body is moving (Schonland's pilot-leader). The discussion above allows us to suggest the occurrence of quantum processes in it. The positive streamer, which spreads from the plasma body back towards the channel, must contain particles, which possess parallel magnetic moments and forming average magnetic field of the streamer and the plasmoid. As the positive streamer spreads from the plasmoid, the increase of the magnetic field in the streamer-plasmoid system must lead to growth of the induction electric field, which should effectively suppress the magnetic field which had created it. The decrease of the average magnetic field should initiate the appearance of an induction electric field of a polarity opposite to the initial one. The moment when a positive streamer connects to the head of the main channel, the current starts to flow along the entire new step, heating it to a high temperature and causing destruction of the stable structure of magnetic moments. When the current flows in the new step, the volume forces in the end area tear it off and through it forward. The arrival of particles with non-compensated magnetic moments into the remaining part of the channel does not happen, and the increase of its length happens, apparently, due to the ionization waves. Then the process repeats.

Differences should be noticed in the development of a positive leader. A group of particles with non-compensated magnetic moments from the leader head penetrates inside the channel, where its magnetic structure is disordered due to the high temperature inside it. Here, as well as for a negative leader, the emerging induction electric field must prevent the growth of a magnetic field. With the decrease of the magnetic field an induction electric field of the opposite polarity should appear. These processes should be less effective for a positive leader due to the ion conductivity in the case of the leader. The magnetic field penetrating inside the channel, creates the surface currents in the conducting lightning channel. Apparently, it is these surface currents that are observed when the currents of the positive leader are measured. If the shape of a positive leader is weakly stepwise then these surface currents should be lowered considerably. The existence in positive leaders of anomalously large currents, compared to the currents of negative leaders, is pointed out by *Carey et al.* [2003], *Bazelyan and Raizer* [2001] and by *Narita et al.* [1989]. *Narita et al.* [1989] presented the data of measured current values of 270 kA in a positive leader of a bipolar flash. *Carey et al.* [2003] give the data from statistic analysis of the peak values of currents in positive and negative leaders. Fig. 5 in the work of *Carey et al.* [2003] shows a histogram of the distribution of the numbers of experimental occurrences of the currents observed in leaders. It can be seen, that in a negative leader the peak values correspond mostly to 20 kA. In positive leaders the peak currents may surpass this value by one order and more. Viewing the bipolar structure of a lightning as a single system of a positive and a negative leader, we can claim that the current-conducting channel is a connecting unit of two separate magnetic systems, each of which contains a streamer zone, a leader head and a transition area. The magnetic system of a negative leader is localized in the plasma object, known as Schonland's pilot leader, which joins briefly the current carrying channel. But how realistic is such a view of a lightning flash, if its consequence is gigantic induction electrical fields of bipolar structure? For a negative leader, the bipolar structure of induction fields is realized in a time interval of new step formation. If an average length of a step is accepted to be 30 m and the speed of the leader

head – $10^6$ m/s, then this time interval corresponds to 30 μs. For the process of lightning development this is a large time interval. Even if we assume that strong induction fields, while spreading, are considerably weakened by the charged disperse environment, their characteristic bipolar structure should be fixed by observers. Apparently, the strong radio-emission of bipolar structure fixed by *Le Vine* [1980], *Willett et al.* [1989], *Smith et al.* [1999] and by *Thomas et al.* [2001] is a proof of the physical scheme of lightning development formulated above. Fig. 4 in the paper of *Smith et al.* [1999] presents for the three time scales the results of measuring of electric field of the negative leader from the distance of 161 km. The length of the bipolar impulse is 25,8±4,9 μs, which corresponds to the time interval of the formation of the new step of the negative leader. In the works of *Smith et al.* [1999] and *Willett et al.* [1989] it is claimed that such impulses should be created by extraordinary physical processes in the lightnings, not known yet. In the work of *Smith et al.* [1999] it is shown that for creation of such impulses in the emission source, the electric fields should be realized with the intensity of up to $2,7 \cdot 10^7$ V/m. These estimates are accepted by *Smith et al.* [1999] as impossible due to lack of experimental evidence.

In the work of *Thomas et al.* [2001] the data are presented from observation of strong radio-emission by lightnings. It is shown that the strongest sources of radio-emissions are emerging in the upper part of a thunderstorm in the area, where the positive charge is located, when the negative leader enters this area. In the lower part of the thunderstorm (in the area of the negative charge) the developing positive leader is a source of radio-emission, the power of which is one order lower than that of a negative leader. The given rule corresponds to the particularities of induction currents in the end areas of the leaders of opposing polarities – ionic and electronic.

The radio-emission, noted by *Thomas et al.* [2001] and by *Le Vine* [1980] is established at the frequencies 3-30 MHz (see [*Smith et al.*, 1999]) and 60 MHz (see [*Thomas et al.*, 2001]). The length of the emission waves corresponds to the spatial scales of the forming new step of the negative leader. The induction fields emerging during the leader development are located in a narrow area perpendicular to the direction of the lightning channel spreading, and should weaken during long distance spreading due to the ionization processes in the media surrounding the lightning channel, the presence of charged particles in the cloud area and the phase distortion on the borders of their spreading area. This is, apparently, the reason why the measuring technology cannot register their large amplitudes. But they are registered with values established by *Le Vine* [1980], *Willett et al.* [1989], *Smith et al.* [1999] and by *Thomas et al.* [2001]. At the same time, the existence of very strong induction fields around the lightning channel may be the reason of its numerous branchings.

The estimate of the value of $B_r$ (relation (3)), given above, is, apparently, close to the real value within one order of magnitude only in the end area of the lightning channel. Outside the channel it should be determined under principally different basic assumptions. And this makes it difficult to receive an estimate of the volume forces, acting upon the end area of the leader. Here it is preferable to have experimental data on the tearing off of the end area of the negative leader. In light of everything noted above, the existence of vortical induction electric fields in the area of the lightning flash spreading appears realistic, which makes also realistic the possibility of appearance, during the process, of the angular momentum of the quantity of motion in the cloud area.

The directions of the induction electric fields in the end areas of a bipolar lightning at the stages of growth and decrease coincide, but the oppositely charged particles, entering the sphere of their action, receive impulses in the opposite direction. For example, when spreading of a negative leader, which enters the area of a positive charge of a cloud system, is considered, along the path of its movement a complex alternate succession of areas with charged particles should appear. These particles possess a tendency to rotational motion in directions corresponding to directions of induction fields at the stages of formation of new steps of a leader. The character of change of a bipolar impulse in Fig. 4 from [*Smith et al.,* 1999] allows suggesting that the process of generation of the rotational motion is determined mostly by the positive part of the impulse, but the final conclusion can only be formulated after careful analysis. A similar picture, apparently, in a weaker variant is realized as well during the spreading of a positive leader.

For a preliminary (within one order of magnitude) estimate of the starting speeds of charged particles, produced by the impulse action on them by the induction fields, the relation (9) can be used and the law of change of a particle impulse.

Thus, on a time interval of the order of 3 μs, corresponding to the main interval of a lifetime of a bipolar radio-impulse (Fig. 4 from [*Smith et al.,* 1999]), at the distance of 10 m from the lightning channel, particles, whose size is between 100 μm and 1 mm, and whose charge is 1 pC, gain the starting speeds between 0,1 m/s and 100 m/s. The impulse gained by the particles is transferred to the surrounding air. Since the presence of water in the cloud media is negligible compared to the air mass in volume unit, the rotational movement, which the environment surrounding the lightning channel might receive, is also negligible. In this connection, the generation of vorticity in the cloud area, sufficient for appearance of an intensive vortices in the atmosphere, may be brought about by the pumping up of the angular momentum by the lightning in the selected volume of the thunderstorm with considerable lightning activity. For a precise definition of what could be the preferable direction of the spinning in the media it is necessary to conduct thorough studies of the processes of generation of induction electric fields during spreading of leaders of different polarities and of distributions of charged particles in the environment and to use the data from observations of intensive vorticity formation. The criteria, linking the size of the area and the lightning activity in it, establishing inevitability of vorticity formation, may be determined by numerical modeling of the phenomenon. Indisputably, this should be based on nature observation data. Thus, for example, *Greneker et al,* [1976] pointed out that by the time of the visible generation of vertical vorticity in the radio-echo image the hole with the diameter of 2 km appeared. If it is assumed that the size of the area, in which the angular momentum from a single lightning is created, is of the order of tens of meters, then several thousands of lightning flashes are required to cover the mentioned area. Such activity is observed during severe storms.

## 4. Conclusion

The formulated suggestion about the existence of an angular momentum of amount of motion in a cloud area, caused by possible appearance of specific distribution of the magnetic field in the end area of the leaders, allows a physically based explanation of a number of particularities of the development of lightning flashes which have not been clear. These include: the stepwise shape of a negative leader, anomalously large currents generated by a positive leader, bursts of extremely strong radio-emissions during the process of lightning development, gamma-ray flashes from storms, and formation of destructive vortices. Also, accepting feasibility of the suggestion is equal to accepting the fact that in a leader head of a lightning channel electromagnetic interaction with Dirak vacuum is occurring, and the birth of electron-positron pairs with their later annihilation. Then, a lightning flash can be identified as a natural gamma-laser. *Vysotskii and Kuzmin* [1989] and *Churikov* [2003] pointed out the possibility of a realization of such a phenomenon based on annihilation of electron-positron pairs. Nevertheless, it is absolutely not clear, what physical process leads to such critical conditions in the local area of a thunderstorm, causing these quantum phenomena. Neither is there clarity about the physical conditions in the proximity of the lightning channel, leading to weakening of strong magnetic and induction electric fields and preventing their registration. Solution to these questions appears as a high-priority problem. It is also necessary to organize observation of lightning flashes development with a temporal resolution, sufficient to trace the formation of a new step of a leader, which may bring about solid arguments in favor of the physical system formulated above.

Conference on meteorological application of lightning data, January, American Meteorological Society, San Diego, CA

Le Vine D. M. (1980), Sources of the Strongest RF Radiation from Lightning, *J. Geophys. Res., 85*(C7), 4091-4095.

Narita K., Y. Goto, H. Komuro and S. Sawada (1989), Bipolar Lightning in Winter at Maki, Japan, *J. Geophys. Res., 94*(D11), 13.191-13.195.

Smith D. A., X. M. Shao, D. N. Holden, C. T. Rhodes, M Brook, P. R. Krehbiel, M. Stanley, W. Rison and R. J. Thomas (1999), A distinct class of isolated intracloud lightning discharges and their associated radio emissions, *J. Geophys. Res., 104*(D4), 4189-4212.

Thomas R.J., P. R. Krehbiel, W. Rison, T. Hamlin, J. Harlin and D. Shown (2001), Observation of VHF Source Powers Radiated by Lightning, *Geophysical Research Letters, 28*(1), 143-146.

Torii T., M. Takeishi and T. Hosono (2002), Observation of gamma-ray dose increase associated with winter thunderstorm and lightning activity, *J. Geophys. Res., 107*(D17), 4324.

Uman M. A. (1969), *Lightning*, McGRAW-HILL Book Company, New-York, St. Louis, San Francisco, Toronto, London, Sydney.

Vonnegut B. (1960), Electrical Theory of Tornadoes, *J. Geophys. Res., 65*(1), 203-212.

Vysotskii V. I. and R. N. Kuzmin (1989), *Gamma-laser,* Moscow State University Publishing House, Moscow.

Willett J. C., J. C. Bailey and E. P. Krider (1989), A Class of Unusual Lightning Electric Field Waveforms With Very Strong High-Frequency Radiation, *J. Geophys. Res.,94*(D13), 16.255-16.267.
19

FIGURE CAPTIONS

Fig. 1. The change of the magnetic induction **B** along the axis of the long solenoid in the end area.

Fig. 2. The scheme of the longitudinal magnetic field **B** and induction electric field $E_\varphi$ in the channel of the bipolar lightning flash. a) positive leader; b) negative leader; 1 – power lines of the electrical field **B**; 2 – the current in the channel; 3 – the induction electric field lines; 4 – the volume force **F**=**I**$_\varphi$×**B**.

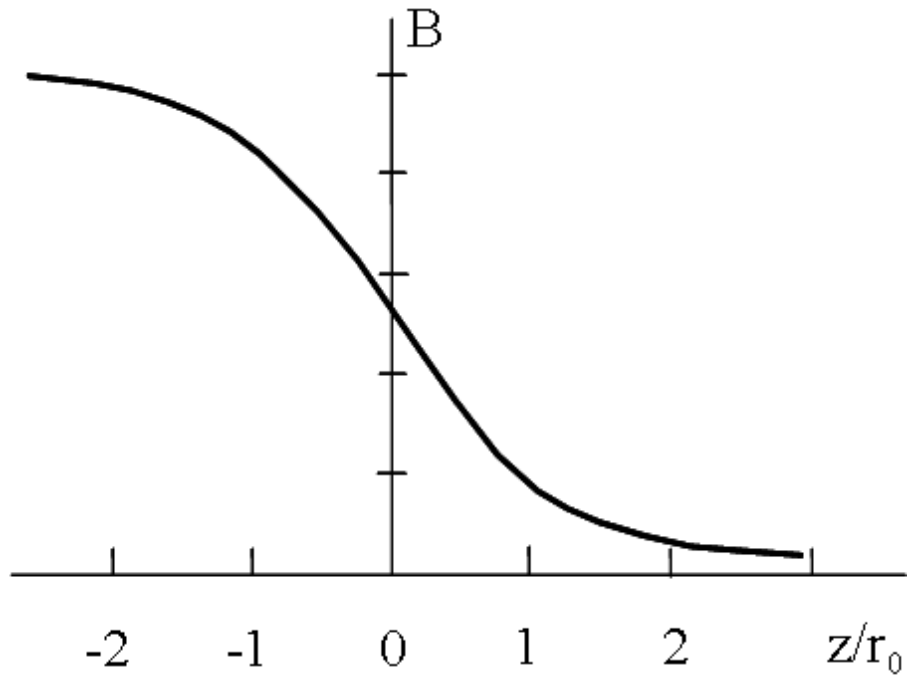

Fig. 1. The change of the magnetic induction $B$ along the axis of the long solenoid in the end area.

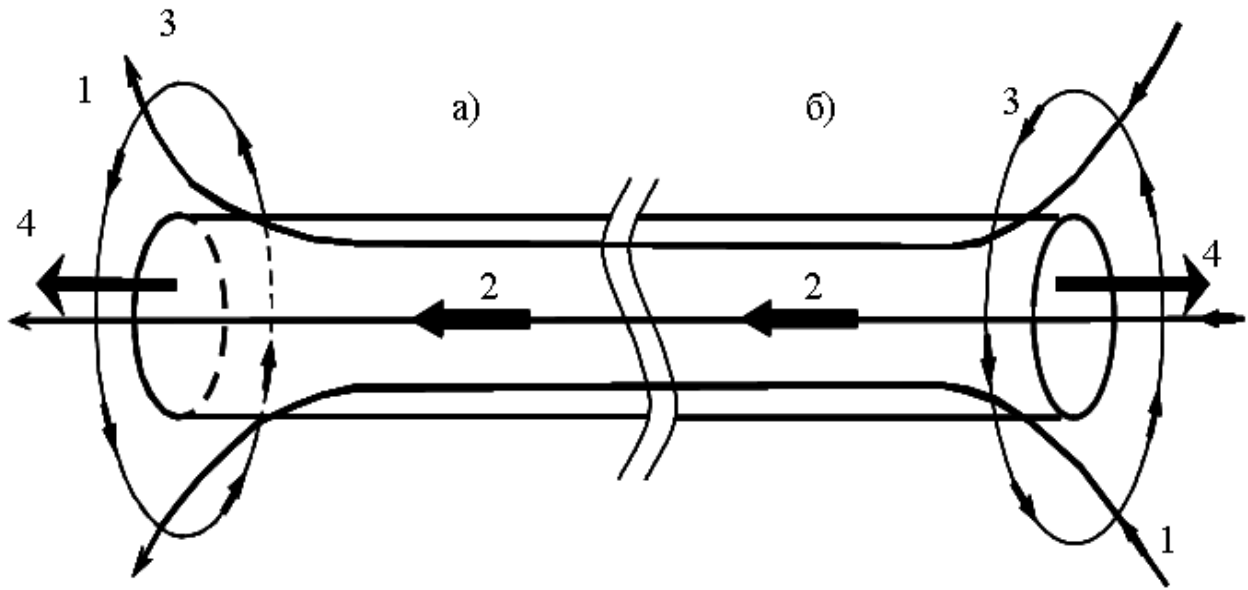

Fig. 2. The scheme of the longitudinal magnetic field **B** and induction electric field $E_\varphi$ in the channel of the bipolar lightning flash. a) positive leader; b) negative leader; 1 – power lines of the electrical field **B**; 2 – the current in the channel; 3 – the induction electric field lines; 4 – the volume force $\boldsymbol{F}=\boldsymbol{I}_\varphi\times\boldsymbol{B}$.